\documentstyle[aps,prb,epsf]{revtex}

\begin{document}

\twocolumn[\hsize\textwidth\columnwidth\hsize\csname
@twocolumnfalse\endcsname

\title{Melting of hard cubes}
\author{E. A. Jagla\cite{autor}}
\address{Centro At\'omico Bariloche\\
Comisi\'on Nacional de Energ\'{\i}a At\'omica\\
8400 S. C. de Bariloche, R\'{\i}o Negro, Argentina}
\maketitle

\begin{abstract}
The melting transition of a system of hard cubes is studied 
numerically both in the case of freely rotating cubes and when there 
is a fixed orientation of the particles $-$parallel cubes. It is 
shown that freelly rotating cubes melt through a first-order 
transition, whereas parallel cubes have a continuous transition 
in which positional order is lost but bond-orientational order remains 
finite. This is interpreted in terms of a defect-mediated
theory of melting
\end{abstract}

\vskip2pc] \narrowtext

\section{Introduction}

Freezing is probably the most unavoidable phase transition of a classical
system of identical particles, occurring when temperature is reduced
sufficiently. The most important difference between the low temperature
(solid) phase and the high temperature (fluid) phase is that in the solid
phase long range correlations between the coordinates of particles exist,
whereas in the fluid phase this correlation is short range, decaying
exponentially with distance. Freezing (or melting) is $-$for three
dimensional systems$-$ a first-order transition, with discontinuities in the
first derivatives of thermodynamic potentials at the transition point.
Compared to the liquid-gas transition or magnetic transitions in model
systems, the advance in the theoretical understanding of the melting
transition has been considerably slow.

Historically, there have been two main theoretical approaches to the problem
of melting. Perturbative arguments\cite{liquidos} starting from the solid
phase are able to predict that the solid structure will be unstable when
temperature is increased sufficiently, but cannot automatically predict the
characteristics of the fluid phase. Descriptions starting from the fluid
phase (virial expansions, Ornstein-Zernicke equations, etc.), although give
very accurate descriptions of the fluid phase in some model systems, usually
do not predict at all a transition to a solid phase. The most successful
account of the melting transition arising from this line of thinking is an
order parameter theory of melting\cite{lry} in which the most stable
structure at a given temperature is obtained by minimizing a free energy
functional which depends on the intensity of the Bragg peaks of the
structure. These intensities are zero in the fluid phase, and finite in the
solid phase.

A more recent approach to the problem is the theory of defect-mediated
melting transition originally proposed for two-dimensional (2D) systems. The
original idea is the following.\cite{hny,kt0} The melting from the solid to
the fluid phase is a consequence of the proliferation of defects in the
perfect crystalline structure that exists at zero temperature. These
defects, called dislocations, appear in pairs defect-antidefect, and have a
finite binding energy. As long as the size of the pair is finite, the
positional order of the system is only perturbed at small length scales.
When temperature is increased beyond a critical value $T_{m}$, the size of
the pair becomes infinite, or in other words, free dislocations can exist in
thermal equilibrium in the system. The existence of free dislocations
disrupts the long range positional order of the system, and drives the
melting transition of the crystal. The transition is predicted to be
continuous, of Kosterlitz-Thouless type.\cite{kt0} It was further realized
that a crystal with a finite concentration of free dislocations does not
behave exactly as a usual fluid. In fact, dislocations destroy the
positional order of crystals, but not the orientational order that can still
be present. In this scenario, the state of the system for $T>T_{m}$
corresponds to a fluid with orientational order, known as the hexatic phase.%
\cite{strand} Within the hexatic phase a new kind of defects appear that
destroy the orientational order at high enough temperatures. These defects
were called disclinations by Halperin and Nelson. At a certain temperature $%
T_{o}$ disclination pairs unbind, and the system transforms in a usual fluid
through a new Kosterlitz-Thouless transition. The critical values $T_{o}$
and $T_{m}$ depend on the self energies and interaction energies of
dislocations and disclinations, which in turn are dependent on the
parameters of the model. It should be noted that a necessary condition for
this two-step melting process is that $T_{o}>T_{m}$, since orientational
order can exist in the absence of positional order, but positional order
cannot exist in the absence of orientational order. If the coupling between
dislocations and disclinations is strong, the two continuous transitions can
merge into a single first-order transition.\cite{ndg,k} Many theoretical and
experimental work was since then devoted to the confirmation of this theory
of two-dimensional melting (that is known as the
Kosterlitz-Thouless-Halperin-Nelson-Young $-$KTHNY$-$ theory) and it was
found in fact that the melting transition in two dimensions may be a single
first order,\cite{2d1o} or a two-step continuous transition\cite{2d2o}
depending on the system studied.

Due to its usefulness in 2D systems, it is tempting to apply the
defect-mediated theory of melting to three-dimensional (3D) systems.\cite
{ndg} In three dimensions, dislocations are one dimensional defects that
form closed loops, or open lines that begin and end at the surfaces of the
sample. In a 3D solid at low temperatures only small dislocation loops
exist, but beyond a critical temperature infinitely large loops are present
at equilibrium and the solid looses its positional long-range order. This
picture of the transition is closely related to the superconducting-normal
transition driven by the proliferation of vortex loops in superconductors,%
\cite{3dxy} or the superfluid-normal liquid transition in He$^{4}$.\cite
{lambda} Again, the transition driven by dislocations alone is continuous
and the fluid above melting still has orientational long range order that
generates a residual resistance to torsion, not present in a normal fluid.%
\cite{nt} In analogy to the situation for 2D systems, a new class of defects
responsible for the disappearance of orientational order must be introduced,
and a possible new transition at higher temperatures should be expected.

From this point of view, the striking feature of the melting of identical 
particles in three dimensions
is that it is always a first-order transition. This may indicate that the
disclination-unbinding and dislocation-unbinding transitions in three
dimensions are strongly coupled, in such a way that they promote each other
and make the transition first-order, but why this is always so is not known.
The existence of a model system that melts through a continuous transition
into a fluid with orientational order would be of importance in giving
insight into the KTHNY theory in three dimensions.

The aim of this work is to analyze numerically a simple model which displays
a continuous melting transition in three dimensions. The model is a system
of impenetrable cubes, which have fixed orientation in space, the same for
all cubes (parallel hard cubes, PHC).\cite{pqphc} Kirkpatrick\cite{kp}
showed that this model has a continuous transition to a simple hypercubic
solid structure in infinite dimensions, and suggested that this would also
be so in three dimensions. The two main reasons to expect a continuous
melting for PHC are the following. First, a cubic crystal lacks $-$in a
Landau description of its melting$-$ a third order term in the free energy
functional that would favor the transition to be first-order.\cite{3g} This
kind of terms appear for crystalline structures that possess three Bragg
vectors $\overrightarrow{G_{1}},$ $\overrightarrow{G_{2}},$ $\overrightarrow{%
G_{3}}$ lying on the first maximum of the diffraction pattern, and
satisfying the relation $\overrightarrow{G_{1}}+\overrightarrow{G_{2}}+%
\overrightarrow{G_{3}}=0$. These vectors do not exist for a simple cubic
structure. In addition, bond-orientational order\cite{boo} will be 
strongly enhanced in PHC
compared for instance to spheres because a fixed orientation of each cube
favors a neighborhood in which cubes arrange with the same orientation. This
raises the possibility for the orientational order to persist up to higher
temperatures than the translational order. For comparison, the case of
freely rotating hard cubes (FRHC) will also be studied, and it will be shown
that in this case the melting is a usual first-order transition into an
isotropic fluid.

\section {Numerical Technique and Results}

The numerical method used to simulate the system is a standard
Monte Carlo-Metropolis algorithm in the $NPT$-ensemble. 
The positions of the cubes are characterized
by the coordinates of their centers. A trial movement of a particle consists
of a displacement to a new position chosen randomly inside a cube of linear
size 0.01 $l$ centered at the old position ($l$ is the linear size of the 
particles). The new position of the particle is accepted as long as there is
no overlap with any other particle. After all particle coordinates are
updated a trial global rescaling of all particle coordinates and system size
by a factor within the range $1\pm .01$ is proposed. If this change does not
produce particle overlapping, then it is accepted according to the
Metropolis algorithm with an energy change $dE$ given by 
$dE=P\Delta V-Nk_BT\Delta V/V$ ($N$ is the total number of
cubes, $P$ is pressure and $V$ is the volume of the system). In
the case of FRHC, in addition to the center-of-particle coordinates, the
three Euler angles are necessary to characterize the position of each cube.
These angles are updated at the same time as coordinates, the elemental
change in each step is chosen to be $\sim 0.1.$ Since there is no 
configurational contribution to the energy of the system, the equation of 
state depends only on the relation $T/P$. All results are presented as 
function of the adimensional temperature $T^* \equiv k_Bv_0^{-1}(T/P)$ 
(that will be refered to simply as ``the temperature"), where $v_0=l^3$ is the
volume of each cube.\cite{notita}

The zero temperature state of the system of cubes (both parallel or freely
rotating) is highly degenerate, because along any of the main crystalline
directions, rows of cubes can be displaced an arbitrary amount without
changing the volume of the system. However, at finite temperatures the cubic
configuration with long range positional order has larger entropy than any
row-displaced configuration, and the thermodynamically stable state is a
simple cubic lattice.\cite{edu} Even for the small systems that we are going
to simulate, this entropy is greater than the one that can be gained by
displacing rows of cubes (which is of the order of $\ln \left( N\right) /N$%
), and configurations with displaced rows never show up in the simulations
in the temperature ranges of interest.

When temperature is increased sufficiently the crystal melts. This melting
is qualitatively different for PHC and for FRHC. In the case of FRHC the
melting occurs via a standard first order transition. Results of simulations
are presented for a system of 125 particles. The system was initialized in a
perfect cubic structure at low temperature, and a simulation was performed
by increasing and then decreasing temperature. At each temperature 5000 Monte 
Carlo steps were used for thermalization and then 20000 
steps were used to compute
quantities of interest. In Fig. \ref{frc1}(a) we see the evolution of the
inverse packing fraction $v \equiv V/(Nv_0)$ of the system. 
It shows a clear hysteretic behavior indicating a
first-order phase transition for $T^*\sim 0.15$, where density changes between 
$\sim 0.45$ and $\sim 0.52$. Also shown in this figure as a dotted line are
the values predicted from a cell-theory for the
solid,\cite{liquidos} which gives a reasonable approximation to the real
equation of state up to the melting temperature. It would be nice to have
the expressions for the virial coefficients of FRHC to fit the fluid part of
the curve, but these are not available to the required order to get a good
fitting. For comparison, in Fig.\ref{frc1} the Carnahan-Starling equation of
state of hard spheres\cite{liquidos} is shown. The only free parameter is
the sphere volume that was chosen to be $1.2 v_0$.

\begin{figure}
\narrowtext
\epsfxsize=3.3truein
\vbox{\hskip 0.05truein
\epsffile{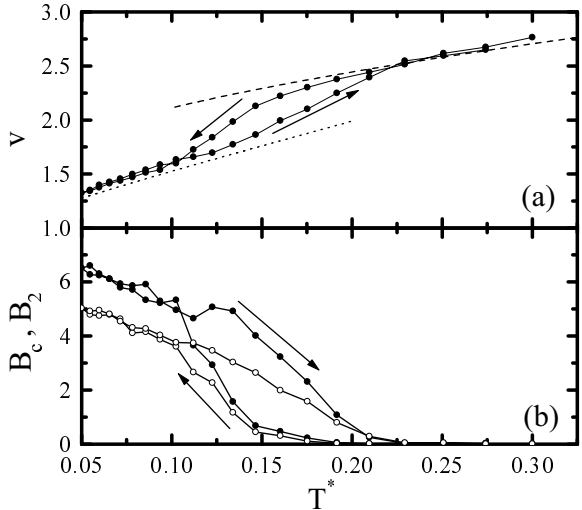}}
\medskip
\caption{Inverse packing fraction $v$ (a), Bragg intensities $B_{c}$ (b, full
symbols) and nearest neighbors orientational order $B_{2}$ (b, open symbols)
as a function of the adimensional temperature $T^*$ 
for a system of $5\times 5\times 5$ FRHC, upon
cooling (starting from a disordered configuration) and heating (see text for
definition, $B_{2}$ and $B_{c}$ are given in arbitrary units). In (a) the
dotted line is the prediction from a cell theory of the solid, and the dashed
line is the behavior of hard spheres with an effective volume of $1.2 v_0$ in 
the fluid phase.}
\label{frc1}
\end{figure}

In Fig.\ref{frc1}(b) two different indicators of the order in the system
support the conclusion that the melting transition of FRHC is first order.
The parameter $B_{c}$ is extracted from the diffraction pattern of the
structure, and it is defined as 
\begin{eqnarray}
B_{c}=\sum_{m=-4}^{4} | \int D\left( k,\theta ,\varphi \right)
Y_{4,m}\left( \theta ,\varphi \right) \times \nonumber \\
\times  \delta \left( k-k_{1}\right)
k^{2}dk\sin \left( \theta \right) d\theta d\varphi | ^{2},
\end{eqnarray}
where $D\left( k,\theta ,\varphi \right) $ is the intensity of the
diffraction pattern in polar coordinates, the delta factor picks up the
values at the first maximum of the diffraction pattern 
($k_{1}=2\pi v ^{1/3}$), and the spherical harmonics $Y_{4,m}$ collect the 
part with cubic symmetry of the diffraction pattern. The value of $B_{c}$ is
different from zero if the system possesses long range positional order.\cite
{notita2}

The relative ordering of neighbor particles $B_{2}$ is defined as 
\begin{eqnarray}
B_{2}=\sum_{m=-4}^{4}| \int D_{2}\left( r,\theta ,\varphi \right)
K\left( r\right) Y_{4,m}\left( \theta ,\varphi \right) \times \nonumber \\
\times  r^{2}dr\sin \left( \theta \right) d\theta d\varphi | ^{2}
\end{eqnarray}
with $D_{2}\left( r,\theta ,\varphi \right) $ being the pair distribution
function of particles at distance $r$, along the spatial direction $\left(
\theta ,\varphi \right) .$ The kernel $K\left( r\right) $ cuts off the
integral beyond some distance. The results are qualitatively insensitive to
the exact form of $K\left( r\right) ,$ in the results presented below $%
K\left( r\right) $ was taken to be 1 for $r<1.5 v ^{1/3},$
and 0 for $r>$ $1.5 v ^{1/3}$. The value of $B_{2}$ is
different from zero if the system possesses long range orientational order.

All these indicators of ordering vanish at the melting transition, with the
same hysteretic behavior as that of the volume. The unambiguous
determination of a first-order phase transition would require the study of
the volume histogram at the transition temperature, which should have a
double peak structure associated to the coexistence of a solid and a fluid
phase. Unfortunately, the simulation of FRHC is very time consuming so as to
carry out this program. Partial checks were performed, however. In a
simulation around the transition temperature ($T^*=.15$) the volume of the
system stabilized around different values, depending if the initial
configuration of the system was chosen random or ordered. These values were
the ones expected from Fig. \ref{frc1}(a). Partial simulations in systems up
to $512$ particles were performed, and the results are consistent with a
first-order melting transition for FRHC.

If the cubes are restricted to be parallel to each other, the nature of the
melting transition changes qualitatively. Results of simulations for this
case are shown in Fig. \ref{phc} for a system of 216 particles. The volume
of the system does not show any abrupt change, but a continuous and
reversible (on heating and cooling) behavior. The parameter characterizing
the crystalline order $B_{c}$ diminishes strongly around $T^*=0.4$, where the
system has a density $\sim 0.5,$ suggesting a continuous melting. The local
orientational order (characterized by $B_{2}$), in spite of decreasing near
the transition remains finite at high temperatures. This characteristic is
not surprising since the orientational order is favored by the equal
orientation of all cubes. In Fig. \ref{phc} we can see also the predictions
for the volume from the lowest order cell model of the solid and the seventh
order virial expansion for the fluid.\cite{7o} These expressions give a good
approximation to the simulated values for all temperatures.

\begin{figure}
\narrowtext
\epsfxsize=3.3truein
\vbox{\hskip 0.05truein
\epsffile{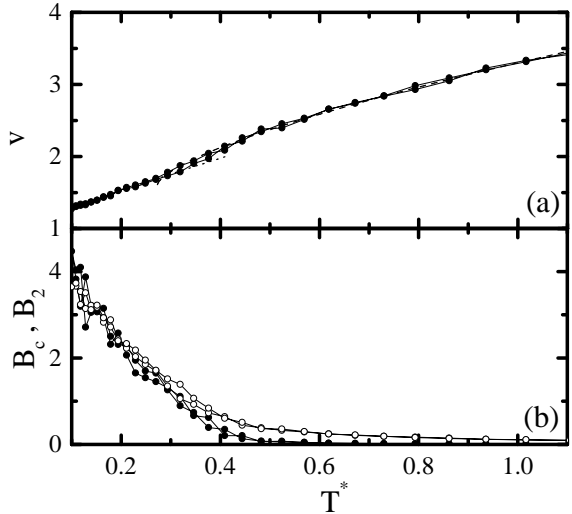}}
\medskip
\caption{Same as Fig.\ref{frc1} for a system of $6\times 6\times 6$ PHC. In
(a) the dotted line is the prediction from a cell theory of the solid, and
the dashed line is the equation of state for PHC to seventh order virial
expansion.}
\label{phc}
\end{figure}

If the melting of PHC is really a second order phase transition, the
behavior of the order parameter of the transition (that can be taken to be the
crystalline order parameter $B_{c}$) must obey scaling laws as a function of
the system size. In particular, different simulations of $B_{c}$ in systems
of different sizes $L$ ($\equiv V^{1/3}/l$) must obey a scaling relation of
the form\cite{staufer} 
\begin{equation}
B_{c}=L^{-\mu }f\left( \left( T^*-T^*_{m}\right) L^{1/\nu }\right) ,
\label{esca}
\end{equation}
where $f$ is a universal function, $\nu $ and $\mu $ are two critical
exponents and $T^*_{m}$ is the thermodynamical melting temperature. The
exponent $\nu $ characterizes the divergence at the thermodynamic melting
temperature $T^*_{m}$ of the correlation length. The result of simulations for
systems of 216, 512, and 1000 particles are shown in Fig. \ref{escaleo}. The
volume and the orientational order $B_{2}$ show no detectable dependence
on size, whereas the crystalline order $B_{c}$ has a clear size dependence.
Results for $B_{c}$ for different system sizes can be collapsed reasonably
well onto a single curve when plotted as $B_{c}L^{\mu }$ vs $\left(
T^*-T^*_{m}\right) L^{1/\nu }$, with parameters $T^*_{m}=0.40\pm 0.02,$ $\mu
=4.0\pm 0.5$, $\nu =0.50\pm 0.05.$ This value of $\nu $ is lower than
the one corresponding to a three dimensional $XY$ model, or the loop
model for the normal-to-superconducting transition ($\nu =.666\pm .003$)
that is supposed to be in the same universality class of our model if the
melting can be described by the KTHNY theory. However, to be able to 
unambiguously decide this point, more simulations in larger systems are needed.
The density of the system at
melting is $0.48\pm 0.02$.

\begin{figure}
\narrowtext
\epsfxsize=3.3truein
\vbox{\hskip 0.05truein
\epsffile{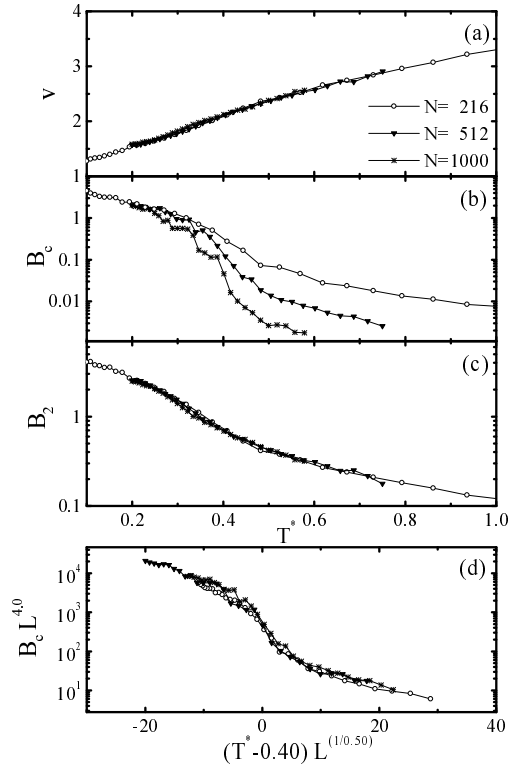}}
\medskip
\caption{Inverse packing fraction $v$ (a), orientational order $B_{2}$ (b), 
and crystalline
order $B_{c}$ (c) as a function of temperature for systems of PHC of
different sizes. The values shown correspond to an average upon cooling and
heating. In (d) the curves of $B_{c}$ are scaled according to a second order
phase transition using the adimensional linear size of the system 
$L \equiv V^{1/3}/l$.}
\label{escaleo}
\end{figure}

The fluid formed by the parallel cubes above melting is not a usual
isotropic fluid. This is obvious since some spatial orientations are singled
out by the particular form of the particles. The fluid phase of PHC is the
analogous of the hexatic phase of the KTHNY theory. Within this framework,
the difference between FRHC and PHC is clear: parallel cubes keep the long
range orientational order even when positional order has been lost, and the
KTHNY theory predicts a continuous melting if only positional order is lost.
FRHC have the possibility of loosing both positional and orientational
order, and this is in fact what happens at a unique temperature in a
discontinuous form. 

It may be of interest to compare the fluid of PHC with the nematic phase of
liquid crystals.\cite{chandra} In that case, molecules orient along a
preferred spatial direction (i.e., they possess molecular-orientational order). 
Upon cooling, this structure transforms usually
into a smectic-A phase in which a long range positional order is established
along the direction characterizing the nematic phase. This transition may be
first or second order depending on the material. At a lower temperature the
smectic-A phase can undergo a transition to a crystalline phase. In our
case, parallel cubes single out three 
orthogonal and equivalent directions in space,
and upon cooling the system freezes into a solid phase, with crystalline
order in all directions. The melting of PHC has no analogous in the
transitions that occur in liquid crystal systems. Note that for the case of
cubes the oriented phase has to be stabilized from outside, whereas the
nematic phase in liquid crystals may be generated by molecular hard core 
interactions only.

\section{Conclusion}

In summary, I have shown numerical results on a simple model that displays a
continuous melting transition in three dimension $-$namely a system of
parallel hard cubes. The melting of this system can be qualitatively 
interpreted in terms
of the KTHNY theory of defect-mediated melting. The melting temperature was
estimated to be $T^*_{m}=0.40\pm 0.02,$ and the critical density is $0.48\pm
0.02.$ The critical exponent of the correlation length is $\nu =0.50\pm 0.05$%
. At the melting transition only positional order is lost, orientational
order remains finite because it is favored by the geometric form of the
particles. If the cubes are allowed to rotate, the melting is a usual first
order transition where both positional and orientational order are lost.

\section{Acknowledgments}

I thank K. Hallberg and D. Dom\'{\i}nguez for critical reading of the 
manuscript. This work was financially supported by Consejo Nacional de 
Investigaciones Cient\'{\i}ficas y T\'{e}cnicas, CONICET, Argentina.


\begin{references}
\bibitem{boo} In the system of PHC, molecular-orientational order, i.e., 
order in the orientation of the individual particles always exist, since
all particles are parallel to each other by definition. Bond-orientational 
order refers to long range correlations in the orientations of the segments 
joining the centers of neighbor particles. This is what is tested with the 
parameter $B_2$ (see text). I will use simply ``orientational order" to refer
to ``bond-orientational order".
\bibitem{notita}  Simulations were also performed by fixing the volume of
the system and calculating the equilibrium pressure (and thus $T^*$) 
along the simulations (the $NVT$-ensemble).
The results are completely consistent with the ones presented here.
\bibitem{notita2}  Strictly speaking, long range orientational order alone
generates values of $B_{c}$ which are different from zero. However, these
values are much smaller than in the presence of long range positional order.
This justifies to use $B_{c}$ as an indicator of long range positional order.
\end{references}
\end{document}